\documentclass[fleqn,10pt]{wlscirep}
\usepackage[utf8]{inputenc}
\usepackage{algorithm}%
\usepackage{algorithmicx}%
\usepackage{algpseudocode}%
\usepackage{hyperref}
\usepackage{nameref}
\usepackage[T1]{fontenc}
\usepackage{multirow}%
\title{DP\textsuperscript{2}FL: Dual Prompt Personalized Federated Learning in Foundation Models}

\author[1]{Ying Chang}
\author[1,2,*]{Xiaohu Shi}
\author[1]{Xiaohui Zhao}
\author[2]{Zhaohuang Chen}
\author[3,*]{Deyin Ma}

\affil[1]{College of Software, Jilin University, Changchun, 130012, China}
\affil[2]{College of Computer Science and Technology, Jilin University, Changchun, 130012, China}
\affil[3]{College of Computer Science and Engineering, Changchun University of Technology, Changchun, 130000, China}

\affil[*]{shixh@jlu.edu.cn; madeyin@ccut.edu.cn}


\begin{abstract}
\textbf{Personalized federated learning (PFL)} has garnered significant attention for its ability to address heterogeneous client data distributions while preserving data privacy. However, when local client data is limited, deep learning models often suffer from insufficient training, leading to suboptimal performance. Foundation models, such as CLIP (Contrastive Language-Image Pretraining), exhibit strong feature extraction capabilities and can alleviate this issue by fine-tuning on limited local data. Despite their potential, foundation models are rarely utilized in federated learning scenarios, and challenges related to integrating new clients remain largely unresolved. To address these challenges, we propose the \textbf{Dual Prompt Personalized Federated Learning (DP\textsuperscript{2}FL)} framework, which introduces dual prompts and an adaptive aggregation strategy. DP\textsuperscript{2}FL combines global task awareness with local data-driven insights, enabling local models to achieve effective generalization while remaining adaptable to specific data distributions. Moreover, DP\textsuperscript{2}FL introduces a global model that enables prediction on new data sources and seamlessly integrates newly added clients without requiring retraining. Experimental results in highly heterogeneous environments validate the effectiveness of DP\textsuperscript{2}FL's prompt design and aggregation strategy, underscoring the advantages of prediction on novel data sources and demonstrating the seamless integration of new clients into the federated learning framework.
\end{abstract}
\begin{document}

\flushbottom
\maketitle
% * <john.hammersley@gmail.com> 2015-02-09T12:07:31.197Z:
%
%  Click the title above to edit the author information and abstract
%
\thispagestyle{empty}

% \noindent Please note: Abbreviations should be introduced at the first mention in the main text – no abbreviations lists. Suggested structure of main text (not enforced) is provided below.

\section*{Introduction}
Recent advancements in deep learning\cite{yang2019federated} have brought remarkable breakthroughs across diverse domains\cite{talaei2023deep}, such as disease diagnosis\cite{avila2024deep,kusumoto2024deep,eskandari2024efficient}, facial recognition\cite{wang2021deep,he2025lmtformer,ma2024face},  video recommendation systems\cite{karatzoglou2017deep,xiang2024integrating}, and emotion recognition\cite{akinpelu2024enhanced,singla2024emotion}. Typically, these methods aggregate all data onto a central server for model training\cite{guendouzi2023systematic}, with model accuracy often strongly correlated with the volume and quality of the data. However, in sensitive fields, centralizing data introduces significant privacy and security challenges\cite{huang2024exploring}.

To mitigate the issue of data silos, Federated Learning (FL)\cite{mcmahan2017communication} has emerged as a promising solution, enabling collaborative model training without direct data sharing. Unlike traditional methods, FL allows a global model to be trained by aggregating parameters from locally trained models on client devices. This approach fundamentally changes the data-handling paradigm, allowing data to remain on clients' devices and only model parameters to be shared with the central server for aggregation.

Classic federated learning models involve a central server and local clients: in each training round, the server distributes the global model to the clients, who train it on their local data and send the updated parameters back to the server for aggregation into a new global model. While approaches like FedAvg\cite{mcmahan2017communication} perform well with similar client data distributions, real-world data is often heterogeneous\cite{gao2022survey}, leading to suboptimal global models. Addressing such data heterogeneity has spurred a new line of research known as Personalized Federated Learning (PFL).

PFL aims to develop personalized models that closely reflect individual clients' data distributions, with strict adherence to data privacy and security requirements\cite{sabah2024model}. PFL can generally be divided into two main categories based on the personalization strategy: Global Model Personalization and Learning Personalized Models\cite{tan2022towards}.

In Global Model Personalization, the focus is on adapting the global federated learning model to individual clients through local adaptation. This approach relies on the generalization capability of the global model, as it directly influences the accuracy of each client's personalized model during local adaptation. To achieve this goal, Duan et al.\cite{duan2020self} proposed Astraea, a framework that addresses label imbalance through Z-score-based data augmentation and downsampling. Additionally, it manages data heterogeneity via a Mediator that reschedules training for clients with skewed data. In contrast to this data-centric approach, FedSteg\cite{yang2020fedsteg} adopts a model-based strategy, wherein transfer learning is utilized to fine-tune the global model for each client after the initial training phase.

In contrast, the Learning Personalized Models approach modifies the aggregation process to directly address clients’ heterogeneous data. A prominent strategy in this category is parameter decoupling. For instance, Arivazhaga et al.\cite{arivazhagan2019federated} divide client models into a base layer, trained globally, and a personalized layer, trained locally. This configuration allows the global layer to capture generalizable features, while the personalized layer reflects each client's unique data distribution. Hanzely et al.\cite{hanzely2020federated} extend this approach by introducing a penalty term to balance model generalization and personalization. Clustering-based approaches have also shown promise; for example, IFCA\cite{ghosh2020efficient} assigns clients to clusters of global models that best suit their data, achieving tailored federated learning.

Despite these advancements, current deep learning frameworks still require large parameter counts, while clients often have limited data, sometimes missing entire classes. Such constraints hinder the adequacy of model training when parameters are aggregated in an FL framework. Large pre-trained Foundation Models, trained on extensive datasets\cite{schneider2024foundation}, offer robust feature extraction capabilities beneficial for various tasks. Fine-tuning these models on small local datasets can yield high-performing models, effectively addressing the problem of insufficient local training data.

Recently, studies like PROMPTFL\cite{guo2023promptfl} have begun integrating foundation models into FL, replacing conventional model training with federated prompt training to reduce parameter requirements. This method outperforms both training from scratch and direct fine-tuning but lacks mechanisms for handling client heterogeneity. Moreover, pFedPrompt\cite{guo2023pfedprompt} leverages the multimodal capabilities of CLIP\cite{radford2021learning} and employs attention mechanisms to effectively capture local client-specific information, thereby enhancing performance in heterogeneous client environments. 

Nevertheless, applications of foundation models in FL are still limited, and existing methods do not address the challenge of integrating new clients dynamically.

To bridge these gaps, we propose Dual Prompt Personalized Federated Learning in Foundation Models (DP\textsuperscript{2}FL), a framework that combines task-awareness with local data-driven insights, effectively leveraging client-specific information captured through prompts to achieve personalized federated learning based on foundation models. DP\textsuperscript{2}FL incorporates two distinct prompts: one that captures federated task information and another that reflects local data distribution. Based on these prompt characteristics, DP\textsuperscript{2}FL employs an aggregation strategy that allows clients to benefit from auxiliary training from other clients while maintaining adaptability to their own data. Furthermore, DP\textsuperscript{2}FL introduces a global model that can make predictions on data from new sources without requiring their participation in the federated learning process. This model also enables the seamless integration of new clients, facilitating efficient onboarding without retraining from scratch.

The core innovations of this work are as follows:
\begin{enumerate}[left=2em]
    \item \textbf{Dual Prompt Design}: In the personalized federated learning framework constructed in this work, a novel dual prompt design is proposed: the task prompt for capturing task-level information, and the data prompt for modeling client-specific data distributions—along with corresponding aggregation strategies.
    \item \textbf{Global Model Adaptation}: A global model designed to extend prediction capabilities to new data sources that have not participated in federated learning training. It also ensures seamless integration of newly added clients without requiring retraining, maintaining both flexibility and efficiency.
\end{enumerate}

\section*{Related Work}
Foundation models, built upon deep neural networks and self-supervised learning\cite{bommasani2021opportunities}, have gained significant attention in recent years due to their robust generalization capabilities. By training on vast, unannotated datasets\cite{wang2023pre}, these models acquire rich semantic knowledge, which enhances their applicability across a wide array of downstream tasks and accelerates the adoption of AI in diverse industries\cite{schneider2024foundation}. Among these models, OpenAI's CLIP, a widely recognized Vision-Language Model (VLM), is distinguished by its effectiveness across diverse tasks. This paper leverages CLIP as the foundation model in our proposed personalized federated learning framework, with a brief introduction to CLIP provided below for context.

As a representative of VLMs, CLIP is pretrained on millions of image-caption pairs, which equips it to simultaneously process textual and visual inputs and learn the semantic relationships between them. Built upon a Transformer\cite{vaswani2017attention} architecture, CLIP's extensive parameters empower it to capture the rich multimodal semantic features essential for a range of applications. However, when applied to domain-specific tasks, CLIP and similar models often encounter limitations due to restricted local training data, resulting in underutilized feature extraction capabilities. To address this challenge, prompt-based learning has emerged as an effective approach, which fine-tunes CLIP's pretrained knowledge to enable more efficient adaptation to specific tasks.

Originally developed within natural language processing (NLP), prompt-based learning guides models to generate task-aligned outputs\cite{brown2020language}. This strategy has since been applied to computer vision and other domains. For foundational models such as CLIP, BERT\cite{devlin2018bert}, and GPT\cite{openai2023gpt4}, a prevalent method involves freezing pretrained parameters while fine-tuning task-specific prompts. This approach enhances task adaptability by capitalizing on the model's existing knowledge while focusing computational resources on refining prompt parameters, thus improving model performance on downstream tasks.

\begin{figure}[ht]
  \centering
  \includegraphics[width=\textwidth]{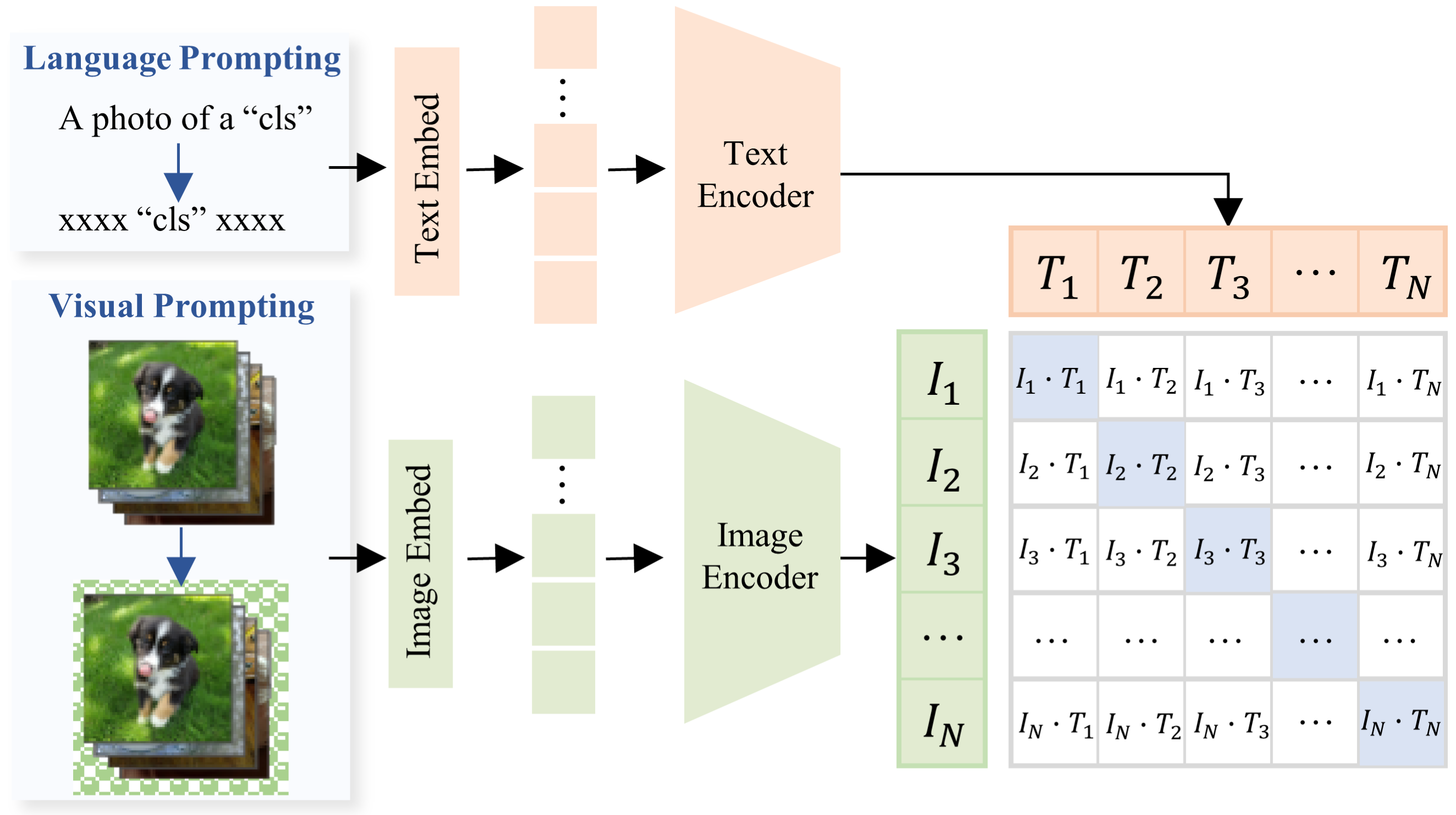}
  \caption{Prompt-Based CLIP Model}
  \label{fig:CLIP}
\end{figure}

As shown in Fig. \ref{fig:CLIP}, research on CLIP-based prompting can be categorized into three primary areas: Language Prompting, Visual Prompting, and Multi-modal Prompting\cite{lei2024prompt}. Language Prompting, focuses on the development of learnable textual contexts within CLIP's text branch to adapt the model for specific downstream tasks. The first work to introduce prompt learning into CLIP was CoOp\cite{zhou2022learning}, which replaced manually crafted prompts with trainable prompt vectors. This shift enabled more efficient adaptation through few-shot learning, significantly reducing training costs. To further enhance generalization, CoCoOp\cite{zhou2022conditional} introduced dynamic adjustments of the trainable prompt vectors in the text branch, using outputs from the image encoder to improve performance across diverse contexts. Recognizing the limitations of a single prompt in capturing both the intrinsic attributes and the extrinsic context of an image, PLOT\cite{chen2022plot} proposed learning multiple prompts collaboratively, leveraging Optimal Transport (OT) to align the visual and textual modalities.

Alternatively, Visual Prompting, as illustrated in Fig. \ref{fig:CLIP}, focuses on modifying the image branch through visual perturbations to improve model training. Bahng et al.\cite{bahng2022exploring} demonstrated the effectiveness of visual prompts for CLIP by exploring three types of prompt applications: random patch insertion, fixed-position patch insertion, and padding. Similarly, ILM-VP\cite{chen2023understanding} explored the influence of label mapping on visual prompting and introduced an automated method for mapping source labels to target labels, which enhanced the accuracy of visual prompts.

While both Language and Visual Prompting modify a single branch of the CLIP model, they do not fully exploit the model's multimodal nature. By contrast, Multi-modal Prompting integrates both Language Prompting and Visual Prompting, allowing the model to simultaneously transform both modalities and thus fully leverage CLIP's inherent multimodal nature. For instance, MaPLe\cite{khattak2023maple} proposed distinct prompts for the text and image branches, which are then coordinated through a coupled adjustment mechanism. This method ensures a high degree of alignment between textual and visual representations, leading to substantial improvements in the model's generalization ability and its adaptability across different domains. The client-side framework utilized in this study builds on the MaPLe architecture.

\section*{Method}

\subsection*{Framework of DP\textsuperscript{2}FL}
This study diverges from traditional personalized federated learning approaches by focusing on adapting federated tasks to foundation models. Given the extensive parameter sizes of foundation models and the typically limited data on federated clients, previous research\cite{guo2023promptfl} demonstrates that training from scratch or parameter fine-tuning often fails to maximize these models' feature extraction capabilities. To address this issue, we propose a prompt-based approach, incorporating a prompt aggregation strategy that optimizes the adaptation of foundation models in federated learning, while minimizing the training parameters required.

\begin{figure}[ht]
  \centering
  \includegraphics[width=\textwidth]{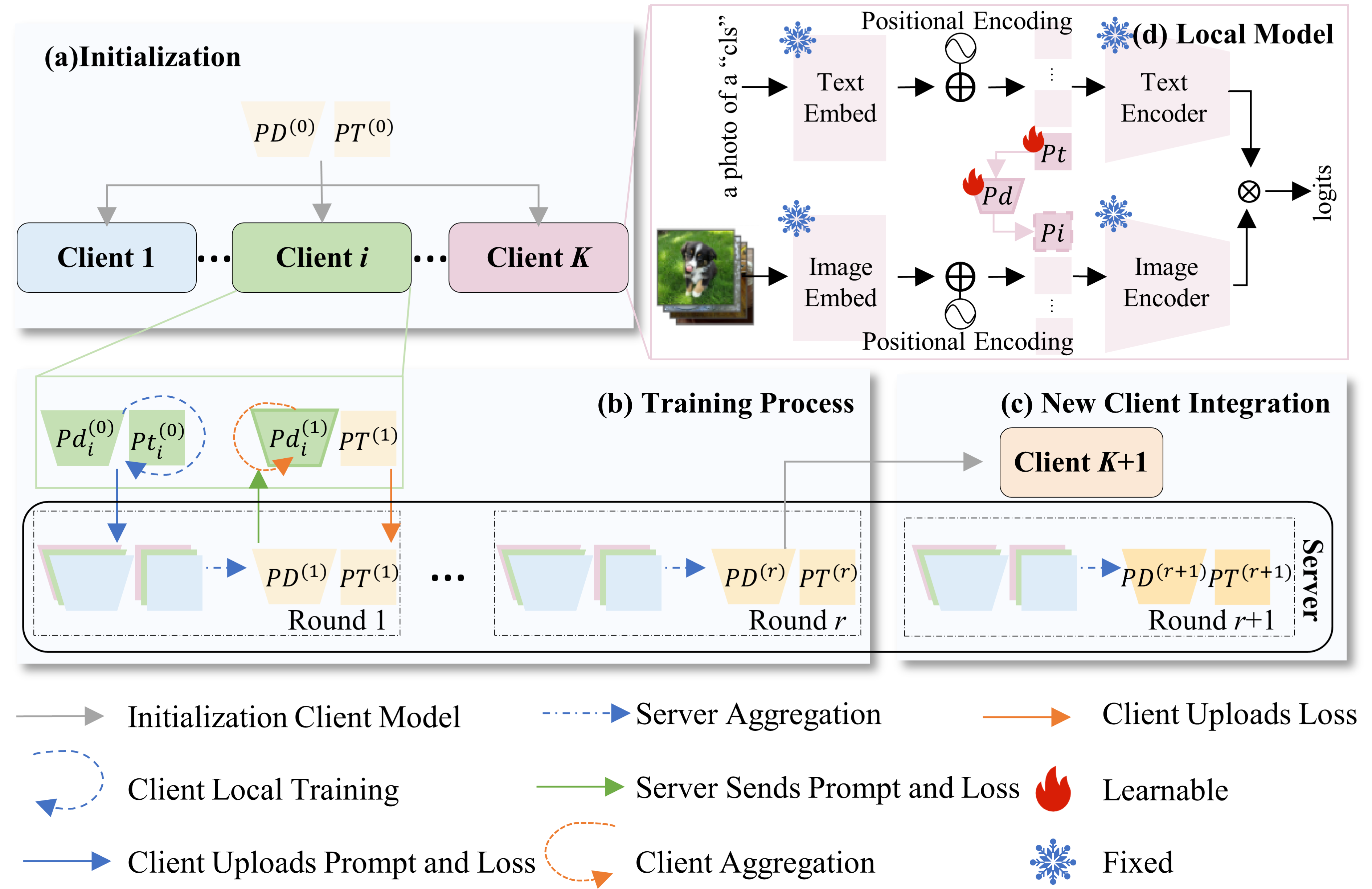}
  \caption{The Framework of DP\textsuperscript{2}FL. The DP\textsuperscript{2}FL workflow comprises three core components: (a) Initialization, which establishes the foundation for federated training; (b) Training Process, which outlines the iterative update and aggregation procedures across clients; (c) New Client Integration, which demonstrates the dynamic onboarding mechanism for new clients. Additionally, (d) Local Model illustrates the client-side framework built upon CLIP.}
  \label{fig:dp2fl_framework}
\end{figure}

In federated learning, the heterogeneity of local data distributions presents a significant challenge in designing universally effective models. To address this, we introduce a dual-prompt strategy that integrates global task alignment with client-specific data characteristics, ensuring that each client's model is effectively tailored to its local data while benefiting from collaborative learning. Furthermore, we propose a global model that facilitates the efficient initialization of newly added clients during training. Traditional initialization methods often incur substantial computational overhead; in contrast, the global model leverages a generalized prompt to streamline client onboarding, allowing new clients to rapidly integrate into the system without the need for extensive retraining. The detailed model framework is illustrated in Fig.\ref{fig:dp2fl_framework}.

The DP\textsuperscript{2}FL framework, as depicted in Fig. \ref{fig:dp2fl_framework}, consists of three critical stages: (a) Initialization, (b) Training Process, and (c) New Client Integration. In the Initialization phase, critical parameters are defined, establishing the foundation for subsequent model training. The Training Process involves iterative refinements of the model, where parameters are updated and aggregated to maintain a balance between generalization and personalization. Finally, the New Client Integration stage tackles the challenge of integrating new clients into the federated task, ensuring their initialization with appropriate parameters, which enables rapid and effective contribution to the learning process. Further details on the CLIP-based Local Model (Fig. \ref{fig:dp2fl_framework}.d) and parameter aggregation strategy are discussed in Sections \hyperlink{sec:Prompt_Design}{Prompt Design} and \hyperlink{sec:Aggregation_Protocol}{Aggregation Protocol}.

\subsubsection*{Initialization}
The Initialization phase begins with the federated task initiator defining essential parameters, such as the model architecture, parameter configuration, and the number of training rounds. Consistent with standard prompt-based learning models, only the prompt components are updated in this framework, while the core parameters of the foundation model remain fixed. The framework incorporates two types of prompts: the task prompt \( Pt \), which captures global task information, and the data prompt \( Pd \), which adapts to each client's specific data distribution. The task prompt is shared among all clients to enable collaborative training through the aggregation of data contributions, while each client maintains a unique local data prompt, with aggregation weights determined by evaluating the relevance of other clients' parameters to the client's local data. This dual-prompt approach ensures the model is tailored to each client's local data while benefiting from collaborative insights.

During this phase, each client uploads a small validation dataset, assumed to be the minimal representative subset of its local data distribution, which is considered shareable for federated learning purposes. The server uses this dataset to compute the initial model loss, which is crucial for guiding the aggregation of model parameters in later stages. As the training progresses, each client evaluates whether the results of other clients' training have improved its model by assessing changes in validation loss. This process guides the parameter aggregation strategy. At the end of the Initialization phase, the server distributes the model parameters, validation data, and loss metrics to all clients, enabling the training process to commence.

\subsubsection*{Training Process}
The Training Process spans \( R \) rounds of federated training. In round \( r \), clients update their prompts by utilizing the previous global task prompt $PT^{(r-1)}$ and their current local data prompt $Pd^{(r-1)}$ as initial model parameters. The updated parameters $\tilde{Pt}^{(r)}$ and $\tilde{Pd}^{(r)}$ are obtained through optimizing a task-relevant loss function on local datasets using stochastic gradient descent (SGD). Subsequently, each client calculates the loss metrics on all validation datasets using its updated parameters and uploads these metrics. The server consolidates the global task prompt $PT^{(r)}$ by evaluating the performance of each client on their validation data and adjusting aggregation weights accordingly. Simultaneously, each client locally adjusts its data prompts $Pd^{(r)}$ by aligning other clients' training outputs with its specific data distribution. After completing these steps, each client computes the loss on its own validation dataset using the updated $PT^{(r)}$ and $Pd^{(r)}$, uploading these losses to the server. These metrics provide essential feedback for the personalized aggregation in the next training round.

Since local data prompts are client-specific, new clients joining the federated learning task must initialize either with the local data prompt parameters established during the initialization phase or with random values, which makes it difficult to align with the existing clients. To address this issue, we introduce a global model composed of both task and data prompts. The task prompt corresponds to the global task prompt described earlier, while the data prompt is generated from the local data prompts using an aggregation method similar to that of the global task prompt. This global model provides a generalized initialization mechanism for new clients, leveraging insights from previous training rounds to enhance adaptability and accelerate their integration into the federated learning framework.

Furthermore, the global model is well-suited for scenarios where new data sources are introduced solely for inference. In such cases, as the data source does not participate in the federated learning process, the global model—comprising both the global task prompt and the global data prompt—can efficiently and directly evaluate the new data.

\subsubsection*{New Client Integration}
The New Client Integration phase is begun when new clients join the federated learning task, either during or after the training process. The new client initially uploads its validation dataset to the server, which distributes it to existing clients to facilitate subsequent aggregation. The server also provides the latest global model for initialization, which includes both the global task prompt $PT$ and the global data prompt $PD$. This global model integrates the training results from all prior rounds, enabling it to demonstrate high accuracy directly on the new client's local dataset. Notably, following initialization, only minimal additional training is needed to adapt the new client's model to its data. Other clients, in turn, integrate the new client's contributions, enhancing the federated model as a whole.

Algorithm \ref{alg:1} presents the complete DP\textsuperscript{2}FL framework process across its three stages.

\begin{algorithm}
\caption{${\rm DP^{2}FL}$ Framework.}
\label{alg:1}
\begin{algorithmic}[1]

\Statex \hspace*{-\algorithmicindent} \textbf{Initialization:}
\State Initialize $PT^{(0)}$, $PD^{(0)}$, $R$
\For{each client $k = 1$ to $K$}
    \State Client $k$ uploads validation datasets $V_k$
    \State Server computes the $loss^{(0)}$ on the $V_k$ using $PT^{(0)}$, $PD^{(0)}$
\EndFor
\State Server sends $PT^{(0)}$, $PD^{(0)}$, $V$, $loss^{(0)}$, $R$ to all clients

\Statex \hspace*{-\algorithmicindent} \textbf{Training Process:}
\For{each round $r = 1$ to $R$}
    \For{each client $k = 1$ to $K$}
        \State Client $k$ utilizing its local dataset to train $PT^{(r - 1)}$, $Pd^{(r-1)}_k$ and obtain $\tilde{Pt}^{(r)}_k$, $\tilde{Pd}^{(r)}_k$
        \State Client $k$ computes the $loss^{(r)}_{k,i}$ on the $V_i \in V$ using $\tilde{Pt}^{(r)}_k$, $\tilde{Pd}^{(r)}_k$
        \State Client $k$ uploads $\tilde{Pt}^{(r)}_k$, $\tilde{Pd}^{(r)}_k$, $loss^{(r)}_{k,i}$
    \EndFor
    \State Server aggregates the $PT^{(r)}$, $PD^{(r)}$ \Comment{Detail in Section \hyperlink{sec:Global_Aggregation}{Global Aggregation}}
    \State Server sends $PT^{(r)}$, $loss^{(r)}$, $Pd^{(r)}$ to all clients
    \State Each client $k$ aggregates $Pd^{(r)}_k$ \Comment{Detail in Section \hyperlink{sec:Local_Aggregation}{Local Aggregation}}
    \State Each client k compute and upload the $loss^{(r)}$ on the $V_k$ using $PD^{(r)}$, $Pd^{(r)}_k$
\EndFor

\Statex \hspace*{-\algorithmicindent} \textbf{Client Addition:}
\State New Client $K + 1$ sends $V_{K + 1}$ to server
\State Server sends $V_{K+1}$ to all clients
\State Server sends the latest global model ($PT$, $PD$) to new client
\State Repeating iterative training
\end{algorithmic}
\end{algorithm}

\subsection*{Prompt Design}
\hypertarget{sec:Prompt_Design}{}

In this study, the CLIP model is leveraged as the foundation for each client's framework, with distinct prompts designed for both the vision and language branches to facilitate cross-modal integration. Specifically, to align visual and textual modalities, a transformation function derives the visual prompt from the textual prompt, as shown in Fig. \ref{fig:dp2fl_framework}.d.

The CLIP model, employing a Vision Transformer (ViT)\cite{dosovitskiy2020image} as its Image Encoder, comprises a sequence of Transformer blocks within both the Text and Image Encoders. In the text branch, for example, the embedded input text, combined with positional encoding, is provided as input to the first Transformer block of the text encoder, represented as follows in Eq. \ref{eq1}:
\begin{equation}
T = Em\_t(t) \oplus 
 En\_t(p)\label{eq1}
\end{equation}
where \( t \) represents the input text, \( p \) denotes the position, \( Em\_t \) refers to the text embedding, and \( En\_t \) is the positional encoding for the text branch, with \( En\_t(p) \in \mathbb{R}^{d\_t} \) matching the dimension of \( Em\_t(t) \).  Here, \( \oplus \) indicates element-wise addition, which is used to combine the semantic information from the text embedding and the positional information from the positional encoding.Similarly, the image branch provides the input to the initial Transformer block of the image encoder as shown in Eq. \ref{eq2}:
\begin{equation}
I = Em\_i(i) \oplus 
 En\_i(p)\label{eq2}
\end{equation}
where \( i \) denotes the input image, and \( Em\_i \) and \( En\_i \) refer to the image embedding and positional encoding for the image branch, respectively, with \( En\_i(p) \in \mathbb{R}^{d\_i} \), matching the dimension of \( Em\_i(i) \).

In this framework, we introduce a task prompt \( Pt \) for the text branch, which is transformed through a dimensional mapping function \( F \) to produce the image prompt \( Pi \) for the image branch, as shown in Eq. \ref{eq3}:
\begin{equation}
Pi = F(Pt)\label{eq3}
\end{equation}
where \( Pi \in \mathbb{R}^{d\_i} \) and \( Pt \in \mathbb{R}^{d\_t} \). Here, \( Pt \) encapsulates the overarching task information within the federated learning setting and is shared uniformly across all clients. Due to the non-identical data distributions typical in federated learning, each client's transformation function \( F \) is adapted via a client-specific parameter set, termed the data prompt \( Pd \), which facilitates personalized adaptation to the client's local dataset.

Under this design, the Text Encoder input in the client model is adjusted from \( T \) to \( [T , Pt] \), while the Image Encoder input is modified from \( I \) to \( [I , Pi] \), preserving the remaining architecture of the CLIP model. During local training on a client's dataset, the parameters of the CLIP model remain fixed, and only the task prompt \( Pt \) and data prompt \( Pd \) are updated.

\subsection*{Aggregation Protocol}
\hypertarget{sec:Aggregation_Protocol}{}
In this study, aggregation strategies are delineated into Global and Local Aggregation based on the participants involved. As discussed in the previous section, the trainable model parameters include two core components: the task prompt, which encapsulates the overarching federated learning task information and is consistent across all clients, and the data prompt, which is tailored to each client, capturing the unique characteristics of local datasets. The task prompt is derived exclusively through a Global Aggregation protocol managed by the server, whereas each client independently computes its data prompt via a Local Aggregation protocol. To enable inference on new data sources and ensure proper initialization for newly added clients, the server computes a generalized data prompt through the Global Aggregation protocol, thereby generating the global model.

\subsubsection*{Global Aggregation}
\hypertarget{sec:Global_Aggregation}{}
To enhance the representation of shared task characteristics in federated learning, the server performs global aggregation on the task prompt. This process, defined in Eq. \ref{eq4}, assigns aggregation weights to client updates based on their performance across all validation datasets, enabling a refined capture of cross-client task information. Following an approach similar to FedFomo\cite{zhang2020personalized}, each client contributes a validation dataset aligned with its local data distribution during initialization, enabling weight assignments proportional to each client's cumulative validation loss.

In the \( r \)-th training round, with \( K \) participating clients, the global task prompt \( PT^{(r)} \) is derived as follows:
\begin{equation}
PT^{(r)}=(\tilde{Pt}^{(r)})^{T} \times W\_T^{(r)}\label{eq4}
\end{equation}
where \( PT^{(r)} \in \mathbb{R}^{d\_t} \) denotes the aggregated global task prompt for round \( r \), subsequently distributed to all clients in round \( r + 1 \) as the initial task prompt for local updates, \( \tilde{Pt}^{(r)} \in \mathbb{R}^{K \times d\_t} \) is the matrix of task prompts trained independently by each client based on local datasets via stochastic gradient descent (SGD), with each row corresponding to a client's task prompt vector. The column vector \( W\_T^{(r)} \in \mathbb{R}^{K} \) contains the aggregation weights for each client in round \( r \), where the weight component \( w\_t_k^{(r)} \) the contribution of client \( k \) to the global task prompt, calculated by:
\begin{equation}
w\_t_k^{(r)}=\frac{ {\textstyle \sum_{i=1}^{K}  {\textstyle \sum_{j=1}^{K}} loss_{i,j}^{(r)}} -  {\textstyle \sum_{j=1}^{K}loss_{k,j}^{(r)}} }{{\textstyle \sum_{i=1}^{K}  {\textstyle \sum_{j=1}^{K}} loss_{i,j}^{(r)}} \times (K-1)} \label{eq5}
\end{equation}
where \( loss_{i,j}^{(r)} \) represents the loss computed by client \( i \) using its round-\( r \) model on the validation set of client \( j \). This weighting scheme reduces the aggregation influence of clients with higher validation losses across datasets, thereby refining \( PT^{(r)} \) to better capture the federated task's overall characteristics.

In addition to aggregating the task prompt, the server uses the Global Aggregation protocol to compute a global data prompt, which provides a generalized representation distinct from the locally optimized data prompts on each client. Together, these form the global model, which enhances adaptability across clients. When new data sources are introduced for inference, or when a new client joins in round \( r + 1 \),  initializing its model parameters with the global model (\( PT^{(r)} \) and \( PD^{(r)} \)) leverages knowledge from prior rounds, thereby reducing the need for extensive retraining. Experimental validation of this initialization effect is presented in Section \hyperlink{sec:Performance_of_Global_Model}{Performance of Global Model}.

\subsubsection*{Local Aggregation}
\hypertarget{sec:Local_Aggregation}{}
In federated learning, clients pursue a common objective despite variations in local data distributions. This framework models the shared task objectives through a task prompt, with the server assigning aggregation weights based on each client's performance across all validation datasets. To address distributional heterogeneity, a data prompt tailored to each client is introduced, enabling local evaluation of models trained by other clients to determine aggregation weights.

In each training round \( r \), the \( K \) clients are divided into three sets based on their contributions to client \( k \): Positive Clients (PC), Retained Negative Clients (RNC), and Discarded Negative Clients (DNC). These sets are formally defined as follows:
\begin{equation}
\text{PC} = \left\{ \text{client}_i \mid loss_{i,k}^{(r)} < loss_{k,k}^{(r-1)},\ 0 < i \leq K \right\}\label{eq6}
\end{equation}
\begin{equation}
\text{RNC} = \left\{ \text{client}_i \mid loss_{k,k}^{(r-1)} \leq loss_{i,k}^{(r)} < \alpha \times loss_{k,k}^{(r-1)},\ 0 < i \leq K \right\}
\label{eq7}
\end{equation}
\begin{equation}
\text{DNC} = \left\{ \text{client}_i \mid loss_{i,k}^{(r)} \geq \alpha \times loss_{k,k}^{(r-1)},\ 0 < i \leq K \right\}
\label{eq8}
\end{equation}
where \( |\text{PC}| + |\text{RNC}| + |\text{DNC}| = K
 \), and \( \alpha \) is the loss tolerance threshold, defined by the task initiator to regulate acceptable performance variations. During aggregation in round \( r \), client \( i \) is classified into PC set if the model loss \( loss_{i,k}^{(r)} \) on client \( k \)'s validation set shows improvement over client \( k \)'s loss from previous round, \( loss_{k,k}^{(r-1)} \). Otherwise, client i is assigned to RNC or DNC based on its performance relative to the defined threshold.

To balance data-specific personalization with generalization, client \( k \) aggregates data prompts from the sets of Positive Clients (PC) and Retained Negative Clients (RNC). The aggregated data prompt \( Pd^{(r)} \) in round \( r \) is computed as follows:
\begin{equation}
{Pd}^{(r)} = {Pd}^{(r-1)} + \text{diag}\left[\left( {W\_d}^{(r)} \right)^T \times \left( \mathbf{1_K}^T \otimes \tilde{Pd}^{(r)} - \mathbf{1_K} \otimes {Pd}^{(r-1)} \right)\right]
\label{eq9}
\end{equation}
where \( {Pd}^{(r)} \in \mathbb{R}^{K \times d\_d} \) represents the locally aggregated data prompt across all clients in round \( r \), and \( \tilde{Pd}^{(r)} \in \mathbb{R}^{K \times d\_d} \) denotes the data prompt computed by each client after training on its dataset. The vector $\mathbf{1_K}$ is a \( K \)-dimensional row vector of ones, and \( \otimes \) denotes the Kronecker product. The matrix \( {W\_d}^{(r)} \in \mathbb{R}^{K \times K} \) represents the aggregation weights, where each element \( {w\_d}^{(r)}_{i,k} \) is the weight of client \( i \) when aggregating client \( k \)'s data prompt. These weights are calculated as follows:
\begin{equation}
{w\_d}_{i,k}^{(r)} = 
\begin{cases}
\text{Norm}(\tilde{w\_d}_{i,k}^{(r)}) & \text{if } \text{client}_i \in \text{PC} \cup \text{RNC} \\
0 & \text{otherwise}
\end{cases}
\label{eq10}
\end{equation}

The initial weight \( \tilde{w\_d}_{i,k}^{(r)} \) is defined as:
\begin{equation}
\tilde{w\_d}_{i,k}^{(r)} = \frac{{loss}_{k,k}^{(r-1)} - {loss}_{i,k}^{(r)}}{\left\| \tilde{Pd}_i^{(r)} - {Pd}_k^{(r-1)} \right\|}
\label{eq11}
\end{equation}
where \( \tilde{Pd}_i^{(r)} \) denotes the data prompt of the model trained on client \( i \)'s dataset after the \( r \)-th round of iteration, and \( {Pd}_k^{(r-1)} \) represents the aggregated data prompt of client \( k \) after the \( (r-1) \)-th round.

From Eq. \ref{eq11}, if \( \text{client}_i \in \text{PC} \), then \( \tilde{w\_d}_{i,k}^{(r)} > 0 \); conversely, \( \text{client}_i \in \text{RNC} \), then \( \tilde{w\_d}_{i,k}^{(r)} \le  0 \). In this work, the normalization function Norm(·) is defined as follows, depending on whether the PC set is empty:
\begin{equation}
\begin{split}
\text{Norm}(\tilde{w\_d}_{i,k}^{(r)}) = \hspace*{26.5em}\\
\begin{cases}
\beta \times \frac{ \sum_{\text{client}_j \in \text{RNC}} \tilde{w\_d}_{j,k}^{(r)} - \tilde{w\_d}_{i,k}^{(r)} }{\sum_{\text{client}_j \in \text{RNC}} \tilde{w\_d}_{j,k}^{(r)} \times (|\text{RNC}| - 1)} & \text{if } |\text{PC}| = 0 \\
\frac{ \tilde{w\_d}_{i,k}^{(r)} \times \text{sgn}(\tilde{w\_d}_{i,k}^{(r)}) + (\gamma \times \tilde{w\_d}_{i,k}^{(r)}) \times (1 - \text{sgn}(\tilde{w\_d}_{i,k}^{(r)})) }{\sum_{\text{client}_j \in \text{PC} \cup \text{RNC}} \left[ \tilde{w\_d}_{j,k}^{(r)} \times \text{sgn}(\tilde{w\_d}_{j,k}^{(r)}) + (\gamma \times \tilde{w\_d}_{j,k}^{(r)}) \times (1 - \text{sgn}(\tilde{w\_d}_{j,k}^{(r)})) \right]} & \text{otherwise}
\end{cases}
\label{eq12}
\end{split}
\end{equation}
where \( \beta \) is a hyperparameter that prevents excessive generalization, \( \gamma \) is an adjustment factor balancing local performance and global generalization, and sgn(·) the sign function defined as:
\begin{equation}
\text{sgn}(x) =
\begin{cases}
    0 & \text{if } x < 0, \\
    1 & \text{otherwise}.
\end{cases}
\label{eq13}
\end{equation}

This aggregation strategy not only utilizes models that perform well on the client's validation set but also considers models within an acceptable error margin, enabling the local model to effectively generalize while adapting to the specific data distribution of the client.

\section*{Experiment}
\subsection*{Experiment Setup}
\subsubsection*{Datasets}
To assess the generalizability of the proposed model framework across diverse data domains, this study evaluates its performance on eight distinct image classification datasets, following the methodologies in\cite{radford2021learning,zhou2022learning,zhou2022conditional}. These include Caltech101\cite{fei2004learning}, which is widely used for general object detection; DTD\cite{cimpoi2014describing}, specialized for texture classification; EuroSAT\cite{helber2019eurosat}, focused on categorizing Sentinel-2 satellite imagery; FGVCAircraft\cite{maji2013fine}, a benchmark for aircraft recognition; Food101\cite{bossard2014food}, tailored for food classification tasks; Flowers102\cite{nilsback2008automated}, dedicated to identifying flower species; OxfordPets\cite{parkhi2012cats}, designed for pet breed classification; and UCF101\cite{soomro2012ucf101}, a leading resource for action recognition studies. Together, these datasets provide a rigorous assessment of the model's cross-domain applicability.

\subsubsection*{Heterogeneity Simulation}
To simulate the non-iid data distributions encountered in real-world federated learning scenarios, we adopt a data-sampling methodology similar to \cite{xu2023personalized,shysheya2022fit}, constructing heterogeneous client datasets that test the framework's ability to leverage foundation models for feature extraction. Following the "16-shot" approach as implemented in CLIP, each client's dataset is constrained to a maximum of 16 samples per category.

For constructing each client's local dataset, we randomly exclude 20\% of the available categories to model data sparsity. Among the remaining categories, 25\% of the data is retained, resulting in a 4-shot structure per category. To further accentuate data heterogeneity, a dominant class is identified randomly for each client; 75\% of the data points in this class are then added to the client's local training set. Both the test and validation datasets are designed to align with each client's training distribution, ensuring consistency across the training, validation, and inference phases. The test dataset is the largest subset that reflects the training distribution, while the validation dataset is the smallest subset. Furthermore, there is no overlap between the datasets used in the training, validation, and inference phases, ensuring mutual exclusivity.

\subsubsection*{Baselines}
Due to the limited research on federated learning with foundation models and the lack of direct comparison methods, we benchmark our framework by integrating traditional personalized federated learning models with the proposed PromptFL \cite{guo2023promptfl}. Specifically, the evaluation includes three baseline models: \textbf{Local}, \textbf{FedProx\cite{li2020federated}+PromptFL(FP+P)}, and \textbf{pFedMe\cite{t2020personalized}+PromptFL(pF+P)}, providing comparative insights into the model's effectiveness across varied federated learning strategies.

\begin{enumerate}
    \item \textbf{Local:} A baseline where each client trains independently on its local data without communication or model aggregation, serving as a non-collaborative reference point.
    \item \textbf{FedProx+PromptFL (FP+P):} FedProx extends the standard FedAvg algorithm by introducing a proximal term to handle heterogeneous data across clients. FedProx+PromptFL combines this approach with the PromptFL method, where clients collaboratively learn task-specific prompts rather than models, enabling federated participants to fine-tune foundation models using minimal local data.
    \item \textbf{pFedMe+PromptFL(pF+P):} pFedMe personalizes federated learning by optimizing both global and local objectives using Moreau envelopes. pFedMe+PromptFL enhances this by learning personalized prompts for each client, allowing for improved adaptation to client-specific data and tasks.
\end{enumerate}

\subsubsection*{Training Details}
Building on the pre-trained ViT-B/16\cite{dosovitskiy2020image} CLIP model, as outlined in MaPLe\cite{khattak2023maple}, the client local model framework in this study is optimized using stochastic gradient descent (SGD) with a learning rate of 0.035. Following the methodology of FedPrompt\cite{zhao2023fedprompt}, the setup includes a centralized server and \( K = 10 \) clients engaged in \( R = 10 \) rounds of iterative training. In each round, clients conduct five epochs with a batch size of four. For local aggregation, parameters are set to \( \alpha = 1.2 \), \( \beta = 0.2 \), with the weight adjustment factor \( \gamma \) defined as follows:
\begin{equation}
\gamma = \frac{1}{3} \times \frac{w\_d_{\min,k}^{(r)}}{w\_d_{\max,k}^{(r)}}
\label{eq14}
\end{equation}
where \( w\_d_{\max,k}^{(r)} \) represents the highest initial aggregation weight among clients in the current RNC set of the \( k \)-th client, while  \( w\_d_{\min,k}^{(r)} \) denotes the lowest initial aggregation weight in the PC set of the \( k \)-th client. All experiments are conducted using PyTorch on an NVIDIA RTX 3090 GPU.

\subsection*{Performance of DP\textsuperscript{2}FL on Heterogeneous Data Distributions}
To validate the efficacy of the proposed parameter aggregation method, comparative experiments were conducted against traditional personalized federated learning approaches, specifically FedProx and pFedMe. Initial results indicate that PromptFL, when applied to training or fine-tuning foundation models from scratch, incurs substantial communication costs and fails to achieve optimal accuracy. Given space constraints, comparisons with scratch training and fine-tuning are omitted. Instead, FedProx and pFedMe frameworks are adapted to integrate PromptFL, restricting training and aggregation to the prompt parameters alone and aligning with the client model structure employed in this work. For quantitative evaluation, the mean accuracy and F1 scores over 10 clients across eight benchmark datasets are reported in Table \ref{tab1}, illustrating the comparative performance of the proposed aggregation strategy.

\begin{table}[ht]
\centering
\begin{tabular}{c|cccc|cccc}
\hline
\multirow{2}{*}{Dataset} & \multicolumn{4}{c|}{ACC} & \multicolumn{4}{c}{Micro-F1} \\
                         & Local & FP+P & pF+P & DP\textsuperscript{2}FL & Local & FP+P & pF+P & DP\textsuperscript{2}FL \\\hline
Caltech101               & 94.51 & \textbf{94.98} & 94.57 & \underline{94.66} & 93.02 & \textbf{94.18} & 93.71 & \underline{93.96} \\
DTD                      & 68.53 & \underline{73.26} & 72.77 & \textbf{73.78} & 64.20 & \underline{70.29} & 70.05 & \textbf{71.14} \\
EuroSAT                  & 78.53 & \underline{83.14} & 81.37 & \textbf{84.18} & 78.16 & \underline{82.99} & 81.22 & \textbf{84.18} \\
FGVCAircraft             & \textbf{43.90} & 41.15 & 41.43 & \underline{41.81} & 34.15 & \underline{37.07} & \textbf{37.17} & 37.06 \\
Food101                  & 87.32 & \underline{88.84} & 88.74 & \textbf{89.06} & 86.05 & \underline{87.92} & 87.78 & \textbf{88.12} \\
Flowers102               & 93.34 & 95.22 & \underline{95.27} & \textbf{95.50} & \underline{91.59} & 94.41 & 94.45 & \textbf{94.91} \\
OxfordPets               & 93.15 & 94.91 & \textbf{95.16} & \underline{94.95} & 92.16 & \underline{94.35} & \textbf{94.61} & 94.21 \\
UCF101                   & 82.47 & \textbf{83.74} & \underline{83.13} & 83.11 & 78.84 & \textbf{82.11} & \underline{81.61} & 81.10 \\\hline
AVG                      & 80.22 & \underline{81.90} & 81.55 & \textbf{82.13} & 77.27 & \underline{80.42} & 80.07 & \textbf{80.59} \\\hline
\end{tabular}
\caption{Comparison Results with Baselines}\label{tab1}
\caption*{\textbf{Bold} values indicate the best performance, and \underline{underlined} values indicate the second best within the same metric across all models. Subsequent tables follow the same convention.}
\end{table}

As shown in Table \ref{tab1}, the proposed framework consistently achieves strong performance across eight datasets, securing the highest accuracy on four datasets and the second-highest on three, resulting in an average accuracy improvement of 0.23\% over the closest competitor. In terms of Micro-F1, DP2FL achieves the highest rank on four datasets and the second-highest on one, with an average improvement of 0.17\% over the second-best method (FP+P). Performance gains are particularly notable on the EuroSAT and Food101 datasets, underscoring the efficacy of the proposed prompt-based aggregation approach. These findings validate the effectiveness of the method in handling heterogeneous client data distributions, surpassing the results of traditional personalized federated learning models and reinforcing its adaptability across diverse data settings.

\subsection*{Performance of Global Model}
\hypertarget{sec:Performance_of_Global_Model}{}
Traditional personalized federated learning methods typically address the variability in local dataset distributions by assigning distinct model parameters to each client. However, these methods fail to account for challenges related to inference on new data sources or the integration of new clients during training, particularly with respect to parameter initialization. To overcome this limitation, we propose a global model that aggregates a generalized data prompt in the same manner as the task prompt. The necessity and effectiveness of this approach are rigorously validated through a series of experiments presented in this section.

Two targeted experiments are conducted to validate the proposed approach. In the first experiment, the average performance of local models trained on the datasets of 10 clients is compared with that of a global model tested on the datasets of all clients. This comparison provides insights into the generalization ability of the global model across different data distributions. In the second experiment, a new client (the 11th client) is introduced. First, the local models trained on the datasets of the initial 10 clients are evaluated on the new data source (the 11th client), and the performance difference between this evaluation and the global model initialization is compared, highlighting the global model’s effectiveness for inference on new data sources. Then, the model is initialized using various methods to demonstrate the role of the global model in initializing new clients. Finally, a round of federated training is conducted, where the newly initialized model is trained alongside the remaining 10 clients. The results show that proper initialization enables the new client to achieve better performance with only a few federated learning iterations.

\textbf{(1) Performance of Global Model on Non-local Datasets.}The global model is pivotal in managing new data sources that cannot directly participate in federated learning. It facilitates inference by enabling accurate predictions on unseen data without requiring retraining. To achieve this, the global model must exhibit strong generalization capabilities, integrating knowledge from multiple clients to effectively address diverse data distributions. Additionally, when new clients join the federated system, the global model should provide an efficient initialization to enable rapid adaptation to local data. Thus, robust generalization is critical not only for inference on unseen data but also for the seamless integration of new clients, enhancing the model’s applicability in real-world federated learning scenarios.

To assess the effectiveness of the global model, a comparative analysis is conducted between the local models and the global model. The results of this comparison are presented in Table \ref{tab4}. In this table, \textbf{Ave\_Local} represents the average accuracy of the local models trained within the DP\textsuperscript{2}FL framework and tested on their respective local datasets. In contrast, \textbf{Global Model} refers to the average accuracy of the global model, initialized with global data and task prompts, and tested across each client’s dataset. The \textbf{Diff} column displays the accuracy difference between the global model and the local models, calculated as the accuracy of the global model (Global Model) minus that of the local models (Ave\_Local).

\begin{table}[ht]
\centering
\begin{tabular}{c|ccc}
\hline
\multicolumn{1}{c|}{{Dataset}} & {Ave\_Local} & {Global Model} & {Diff} \\ \hline
Caltech101    & 94.66 & 94.29 & -0.37  \\
DTD           & 73.78 & 73.52 & -0.26  \\
EuroSAT       & 84.18 & 83.94 & -0.24  \\
FGVCAircraft  & 41.81 & 41.17 & -0.64  \\
Food101       & 89.06 & 88.91 & -0.15  \\
Flowers102    & 95.50 & 94.80 & -0.70  \\
OxfordPets    & 94.95 & 94.86 & -0.09  \\
UCF101        & 83.11 & 82.89 & -0.22  \\
\hline
AVG  & 82.13 & 81.80 & -0.33  \\ \hline
\end{tabular}
\caption{Performance of Global Model on Non-local Datasets}\label{tab4}
\end{table}

As shown in Table \ref{tab4}, the results indicate that the global model performs slightly worse than the locally personalized models across all datasets. The average accuracy difference between the global model (Global Model) and the local models (Ave\_Local) is -0.33\%, reflecting a marginal decline in performance when using the global model. Notably, the global model's accuracy closely aligns with that of the local models on datasets such as OxfordPets and Food101, with minimal differences of -0.09\% and -0.15\%, respectively.

The experimental results suggest that the global model demonstrates strong generalization capabilities, enabling effective inference on new, unseen data sources and efficient initialization of newly added clients.

\textbf{(2) Performance of the Global Model in New Client Initialization.} Building on the findings from the first experiment, which demonstrated the global model's strong generalization ability, this experiment further investigates its performance when new clients are introduced. Specifically, it evaluates the model's inference ability on new data sources and examines how different initialization strategies affect the performance of newly added clients in the federated learning process. The experiment consists of three parts: First, it compares the global model’s inference performance on new data sources with that of locally trained models from other clients. Second, it applies various initialization strategies to the newly added client’s model, assessing their impact on performance with the client’s local dataset. Finally, the results show that after effectively initializing the new client’s model, only a minimal number of federated learning iterations are required to achieve strong performance across diverse datasets.

Table \ref{tab2} presents the results for the first two parts of this experiment. The methods are described as follows: \textbf{Ave\_Local} represents the average accuracy when the original 10 clients perform inference on the new client’s dataset using their locally trained models. \textbf{Global Model} uses the global model, initialized with both task and data prompts aggregated through global aggregation, specifically \( PT^{(R)} \) and \( PD^{(R)} \)and then performs inference to obtain the resulting accuracy. This forms the first part of the comparison. In the second part, additional initialization methods are explored: \textbf{Init} initializes the new client’s model with parameters set by the task initiator, specifically \( PT^{(0)} \) and \( PD^{(0)} \); \textbf{InitGlo} initializes the task prompt using global parameters based on the aggregation method, while the data prompt is initialized with parameters from the task initiator, i.e., \( PT^{(R)} \) and \( PD^{(0)} \).

\begin{table}[ht]
\centering
\begin{tabular}{@{}c|cccc@{}}
\hline
Dataset     & Ave\_Local & Global Model & Init & InitGlo  \\ \hline
Caltech101  & 97.37      & \textbf{97.84}        & 95.69 & 96.55  \\
DTD         & 67.12      & \textbf{67.14}        & 43.03 & 63.83  \\
EuroSAT     & \textbf{84.00}      & 83.85        & 45.21 & 78.35  \\
FGVCAircraft& 41.56      & \textbf{41.90}        & 26.07 & 41.31  \\
Food101     & 89.14      & \textbf{89.15}        & 87.36 & 88.91  \\
Flowers102  & \textbf{92.58}      & 91.82        & 69.50 & 90.88  \\
OxfordPets  & 95.76      & \textbf{96.10}        & 87.79 & 95.90  \\
UCF101      & 82.73      & \textbf{82.91}        & 63.12 & 79.27  \\ \hline
AVG         & 81.28      & \textbf{81.34}        & 64.72 & 79.38  \\ \hline
\end{tabular}
\caption{Results for New Clients under Different Parameter Initialization Methods without Training}\label{tab2}%
\end{table}

% \begin{table}[h]
% \captionsetup{width=\textwidth}
% \caption{Results for New Clients under Different Parameter Initialization Methods without Training}\label{tab2}
% % \resizebox{\textwidth}{!}{
% \begin{tabular}{c|cccc}
% \hline
% Dataset     & Ave\_Local & Global Model & Init & InitGlo  \\ \hline
% Caltech101  & 97.37      & \textbf{97.84}        & 95.69 & 96.55  \\
% DTD         & 67.12      & \textbf{67.14}        & 43.03 & 63.83  \\
% EuroSAT     & \textbf{84.00}      & 83.85        & 45.21 & 78.35  \\
% FGVCAircraft& 41.56      & \textbf{41.90}        & 26.07 & 41.31  \\
% Food101     & 89.14      & \textbf{89.15}        & 87.36 & 88.91  \\
% Flowers102  & \textbf{92.58}      & 91.82        & 69.50 & 90.88  \\
% OxfordPets  & 95.76      & \textbf{96.10}        & 87.79 & 95.90  \\
% UCF101      & 82.73      & \textbf{82.91}        & 63.12 & 79.27  \\ \hline
% AVG         & 81.28      & \textbf{81.34}        & 64.72 & 79.38  \\ \hline
% \end{tabular}
% % }
% \end{table}

In the first part of the experiment, the 11th client is treated as a new data source. Without the global model, each client would likely have to rely solely on its locally trained model to handle the new data. To evaluate the effectiveness of the global model for inference on new data, we compared it with the Ave\_Local method. As shown in Table \ref{tab2} (first and second columns), the global model generally outperforms the average of the local models across most datasets. For instance, on the Caltech101 dataset, the global model achieves an accuracy of 97.84\%, which is higher than the 97.37\% of the average local model. Similarly, on OxfordPets, the global model reaches 96.10\%, compared to 95.76\% from the local models. In several other datasets, such as DTD and Food101, the global model shows an improvement over the average local model, although the difference is not always substantial.

This comparison confirms the advantage of using the global model for inference on new data sources, demonstrating its value in federated learning systems and validating its capability to handle new data sources effectively.

The second part of the experiment evaluates the effectiveness of the global model for new client integration, as shown in the last three columns of Table \ref{tab2}. The Global Model consistently outperforms both Init and InitGlo, achieving the highest accuracy across all datasets. Specifically, the global model improves accuracy by an average of 1.96\% compared to InitGlo, and by 16.62\% compared to Init.

The results from this part highlight the effectiveness of the global model in initializing new clients within federated learning systems, demonstrating its ability to facilitate direct adaptation to the data distribution of newly added clients.

To further validate that proper initialization of the newly added client leads to good performance with fewer subsequent federated learning iterations, this section presents the third part of the experiment. As shown in Table \ref{tab3}, three distinct test metrics are considered: \textbf{New}, which represents the accuracy of the newly added (11th) client’s model on its own local dataset after one round of federated training; \textbf{Local}, which refers to the average accuracy of all 11 clients evaluated on their respective local datasets; and \textbf{All}, which indicates the average accuracy of all 11 clients, where each client’s model is evaluated on the combined test sets using their individual parameters.

\begin{table}[ht]
\centering
\begin{tabular}{c|ccc|ccc|ccc}
\hline
\multirow{2}{*}{Dataset} & \multicolumn{3}{c|}{New}                         & \multicolumn{3}{c|}{Local}                       & \multicolumn{3}{c}{All}                          \\ 
                         & Init           & InitGlo        & Global Model           & Init           & InitGlo        & Global Model           & Init           & InitGlo        & Global Model           \\ \hline
Caltech101               & \textbf{97.84} & \textbf{97.84} & \textbf{97.84} & \textbf{94.85} & \underline{94.84}    & \underline{94.84}    & 95.17          & \underline{95.29}    & \textbf{95.31} \\
DTD                      & \underline{67.61}    & 66.43          & \textbf{69.74} & 72.68          & \underline{72.81}    & \textbf{73.42} & 67.22          & \underline{67.39}    & \textbf{67.61} \\
EuroSAT                  & 82.13          & \underline{82.31}    & \textbf{83.27} & 83.00          & \underline{83.58}    & \textbf{84.14} & 80.54          & \underline{81.14}    & \textbf{81.31} \\
FGVCAircraft             & 42.26          & \underline{43.69}    & \textbf{44.40} & 41.86          & \textbf{42.67} & \underline{42.38}    & 35.56          & \textbf{36.01} & \underline{36.00}    \\
Food101                  & \underline{89.36}    & 89.26          & \textbf{89.47} & \underline{89.09}    & \textbf{89.10} & 89.05          & \textbf{87.36} & \underline{87.34}    & 87.29          \\
Flowers102               & \underline{92.77}    & \textbf{93.40} & 92.45          & 94.89          & \underline{95.41}    & \textbf{95.47} & 92.72          & \underline{93.57}    & \textbf{93.59} \\
OxfordPets               & \textbf{95.79} & \underline{95.69}    & \underline{95.69}    & \textbf{94.90} & 94.83          & \underline{94.87}    & \textbf{93.50} & \textbf{93.50} & \underline{93.47}    \\
UCF101                   & \textbf{83.58} & 82.50          & \underline{83.31}    & \textbf{83.11} & 83.03          & \underline{83.06}    & 80.44          & \underline{80.54}    & \textbf{80.56} \\ \hline
AVG                      & \underline{81.42}    & 81.39          & \textbf{82.02} & 81.80          & \underline{82.04}    & \textbf{82.15} & 79.06          & \underline{79.35}    & \textbf{79.39} \\ \hline
\end{tabular}
\caption{Comparison Results on Different Test Data after One Round Training}\label{tab3}
\end{table}

The results presented in Table \ref{tab3} demonstrate the efficacy of the proposed initialization method, which significantly facilitates the adaptation of the newly added client to various data distribution with minimal training. Notably, the Global Model initialization method consistently outperforms the other initialization strategies (Init and InitGlo) across all performance metrics: New, Local, and All. 

In particular, the newly introduced client achieves the highest accuracy on its local test set with the Global Model across most datasets, as shown in the \textbf{New} column of Table \ref{tab3}. While there are a few cases (e.g., Caltech101) where Init and InitGlo show similar performance, the Global Model consistently yields the best average results. This demonstrates its effectiveness in helping the new client quickly adapt to its local data distribution.

Regarding the performance of existing clients, the \textbf{Local} column indicates that Global Model also leads to the highest average accuracy for the models of all 11 clients. In particular, datasets like DTD, EuroSAT, and Flowers102 show significant improvements with Global Model, highlighting the benefit of this initialization strategy in enabling existing clients to better utilize the new data introduced by the added client.

Finally, when evaluating the aggregated performance across all clients (the \textbf{All} column), Global Model consistently outperforms the other initialization methods. Compared to Init and InitGlo, Global Model demonstrates a clear advantage, particularly on datasets such as DTD and Flowers102. By achieving the highest accuracy on the combined test sets, it showcases its superior generalization ability across diverse data distributions, making it the most robust initialization strategy for federated learning scenarios.

In summary, the third part of the experiment demonstrates that by initializing new clients with the global model, only a few rounds of federated training are required to achieve strong performance across various datasets. Together, the three experiments further validate the global model's robust generalization ability, underscoring its potential for inference on new data sources and for efficiently initializing newly added clients.

\section*{Conclusion}
Federated learning in heterogeneous environments presents significant challenges, primarily due to the substantial variation in local data distributions and the frequent addition of new clients. To address these challenges, we propose the Dual Prompt Personalized Federated Learning (DP\textsuperscript{2}FL) framework. This framework leverages a dual-prompt mechanism and adaptive aggregation strategies to effectively integrate global task information with client-specific data. It enhances the model's generalization to global tasks while accommodating the unique characteristics of each client's local distribution.

A key innovation of DP\textsuperscript{2}FL is the introduction of a novel global model, which enables high-accuracy inference on new data sources that have not participated in federated learning. It also facilitates the seamless integration of new clients into the federated learning process. Empirical results demonstrate that DP\textsuperscript{2}FL enhances model performance across diverse client distributions, improves inference accuracy on unseen data, and reduces the onboarding time for new clients, thereby increasing its practical applicability.

\section*{Data availability}The datasets used in this study are all publicly available and widely adopted in the research community. Specifically, the following datasets were utilized: Caltech101 \cite{fei2004learning} for general object detection, DTD \cite{cimpoi2014describing} for texture classification, EuroSAT \cite{helber2019eurosat} for Sentinel-2 satellite imagery categorization, FGVCAircraft \cite{maji2013fine} for aircraft recognition, Food101 \cite{bossard2014food} for food classification tasks, Flowers102 \cite{nilsback2008automated} for flower species identification, OxfordPets \cite{parkhi2012cats} for pet breed classification, and UCF101 \cite{soomro2012ucf101} for action recognition studies. These datasets are accessible through their respective repositories:  
\begin{itemize}
    \item \textbf{Caltech101}: \url{http://www.vision.caltech.edu/Image_Datasets/Caltech101/}
    \item \textbf{DTD}: \url{https://www.robots.ox.ac.uk/~vgg/data/dtd/}
    \item \textbf{EuroSAT}: \url{http://madm.dfki.de/files/sentinel/EuroSAT.zip}
    \item \textbf{FGVCAircraft}: \url{http://www.robots.ox.ac.uk/~vgg/data/fgvc-aircraft/}
    \item \textbf{Food101}: \url{ https://data.vision.ee.ethz.ch/cvl/datasets_extra/food-101/}
    \item \textbf{Flowers102}: \url{http://www.robots.ox.ac.uk/~vgg/data/flowers/102/}
    \item \textbf{OxfordPets}: \url{https://www.robots.ox.ac.uk/~vgg/data/pets/}
    \item \textbf{UCF101}: \url{https://drive.google.com/file/d/10Jqome3vtUA2keJkNanAiFpgbyC9Hc2O/view}
\end{itemize}

These datasets ensure reproducibility of the experiments and facilitate further research.

\bibliography{sn-bibliography}

\begin{thebibliography}{99}

\bibitem{yang2019federated} Yang, Qiang and Liu, Yang and Chen, Tianjian and Tong, Yongxin. \emph{Federated machine learning: Concept and applications}. ACM Transactions on Intelligent Systems and Technology (TIST), 10(2):1--19, 2019.

\bibitem{talaei2023deep} Talaei Khoei, Tala and Ould Slimane, Hadjar and Kaabouch, Naima. \emph{Deep learning: Systematic review, models, challenges, and research directions}. Neural Computing and Applications, 35(31):23103--23124, 2023.

\bibitem{he2025lmtformer} He, Lang and Zhao, Junnan and Zhang, Jie and Jiang, Jiewei and Qi, Senqing and Wang, Zhongmin and Wu, Di. \emph{LMTformer: facial depression recognition with lightweight multi-scale transformer from videos}. Applied Intelligence, 55(2):195, 2025.

\bibitem{sun2025machine} Sun, Qi and Cheng, Xin and Han, Kuo and Sun, Yichao and Ren, He and Li, Ping. \emph{Machine learning-based assessment of diabetes risk}. Applied Intelligence, 55(2):1--13, 2025.

\bibitem{sharma2022deep} Sharma, Deepak Kumar and Chatterjee, Mayukh and Kaur, Gurmehak and Vavilala, Suchitra. \emph{Deep learning applications for disease diagnosis}. In Deep learning for medical applications with unique data, pages 31--51. Elsevier, 2022.

\bibitem{khan2021machine} Khan, Protima and Kader, Md Fazlul and Islam, SM Riazul and Rahman, Aisha B and Kamal, Md Shahriar and Toha, Masbah Uddin and Kwak, Kyung-Sup. \emph{Machine learning and deep learning approaches for brain disease diagnosis: principles and recent advances}. IEEE Access, 9:37622--37655, 2021.

\bibitem{avila2024deep} '{A}vila-Jim'{e}nez, Jos'{e} Luis and Cant'{o}n-Habas, Vanesa and Carrera-Gonz'{a}lez, Mar'{\i}a del Pilar and Rich-Ruiz, Manuel and Ventura, Sebasti'{a}n. \emph{A deep learning model for Alzheimer’s disease diagnosis based on patient clinical records}. Computers in Biology and Medicine, 169:107814, 2024.

\bibitem{ma2024face} Ma, Jianxiang. \emph{Face recognition technology and privacy protection methods based on deep learning}. In International Conference on Computer Application and Information Security (ICCAIS 2023), volume 13090, pages 899--904. SPIE, 2024.

\bibitem{wang2021deep} Wang, Mei and Deng, Weihong. \emph{Deep face recognition: A survey}. Neurocomputing, 429:215--244, 2021.

\bibitem{karatzoglou2017deep} Karatzoglou, Alexandros and Hidasi, Bal'{a}zs. \emph{Deep learning for recommender systems}. In Proceedings of the Eleventh ACM Conference on Recommender Systems, pages 396--397, 2017.

\bibitem{xiang2024integrating} Xiang, Yafei and Huo, Shuning and Wu, Yichao and Gong, Yulu and Zhu, Mengran. \emph{Integrating AI for Enhanced Exploration of Video Recommendation Algorithm via Improved Collaborative Filtering}. Journal of Theory and Practice of Engineering Science, 4(2):83--90, 2024.

\bibitem{guendouzi2023systematic} Guendouzi, Badra Souhila and Ouchani, Samir and Assaad, Hiba EL and Zaher, Madeleine EL. \emph{A systematic review of federated learning: Challenges, aggregation methods, and development tools}. Journal of Network and Computer Applications, 2023.

\end{thebibliography}

\section*{Acknowledgements}
This work was supported by the Science-Technology Development Plan Project of Jilin Province (20210202129NC, 20230201073GX). The author acknowledges with thanks the support from Jilin Province for both technical and financial assistance.

\section*{Author contributions}
Ying Chang: Conceptualization, Methodology, Software, Visualization, Writing-original draft. Xiaohu Shi: Supervision, Conceptualization, Writing-review. Xiaohui Zhao: Writing-review. Zhaohuang Chen: Writing-review. Deyin Ma: Supervision, Writing-review.

\section*{Funding}
This work was funded by the Science-Technology Development Plan Project of Jilin Province (20210202129NC, 20230201073GX).

\section*{Declarations}
\subsection*{Competing interests}The authors declare that there are no known competing financial interests or personal relationships that could have appeared to influence the work reported in this paper.

\end{document}